%% file: main.tex
\documentclass[10pt,twocolumn,letterpaper]{article}

\usepackage{cvpr}              

\usepackage{graphicx}
\usepackage{amsmath}
\usepackage{amssymb}
\usepackage{booktabs}
\usepackage{times}
\usepackage{epsfig}
\usepackage{graphicx}
\usepackage{amsmath}
\usepackage{amssymb}
\usepackage{booktabs}
\usepackage{multirow}
\usepackage{array}
\usepackage{mathrsfs}
\usepackage{float}
\usepackage[skip=1pt]{subcaption}
\usepackage{xcolor}
\usepackage[skip=1pt]{caption}
\usepackage[export]{adjustbox}

\newcolumntype{C}[1]{>{\centering\arraybackslash}p{#1}}

\usepackage[pagebackref,breaklinks,colorlinks]{hyperref}
\usepackage[symbol,hang,flushmargin]{footmisc} 

\usepackage[capitalize]{cleveref}
\crefname{section}{Sec.}{Secs.}
\Crefname{section}{Section}{Sections}
\Crefname{table}{Table}{Tables}
\crefname{table}{Tab.}{Tabs.}

\newcommand{\klens}{K\textbar Lens GmbH, Germany}
\newcommand{\uni}{Saarland Informatics Campus, Germany}
\newcommand{\dfki}{German Research Center for Artificial Intelligence (DFKI), Germany}
\newcommand{\mpi}{MPI Informatics, Germany}
\newcommand{\iitj}{IIT Jammu, India}

\makeatletter
\def\thanks#1{\protected@xdef\@thanks{\@thanks
		\protect\footnotetext{#1}}}

\def\cvprPaperID{22} 
\def\confName{CVPR}
\def\confYear{2023}

\usepackage{cuted}
\usepackage{capt-of}

\begin{document}

\title{\vspace{-1.5cm}Expanding Synthetic Real-World Degradations for Blind Video Super Resolution}
\vspace{-2cm}
	\author{Mehran Jeelani$^{1,2}$$^{,}$*, Sadbhawna$^{4,}$*,  Noshaba Cheema$^{2,3,5}$, Klaus Illgner-Fehns$^{1}$, \\ Philipp Slusallek$^{2,5}$, and Sunil Jaiswal$^{1,\dag}$
		\vspace{0.5cm} \\
		$^{1}$ \klens,	$^{2}$ \uni, $^{3}$ \mpi\\
		$^{4}$ \iitj, $^{5}$ \dfki   \\
		\thanks{		\hspace{-0.6cm} This work was partially funded by the German Ministry for Education and Research
			(BMBF) under the grant PLIMASC and Indo-German Science and Technology Center (IGSTC).}
	}
	\maketitle
	\footnotetext[1]{Both authors contributed equally to the paper}
	\footnotetext[2]{Corresponding author: {\tt\small sunil.jaiswal@k-lens.de}}
\input{figures/cover.tex}

\begin{abstract}
Video super-resolution (VSR) techniques, especially deep-learning-based algorithms, have drastically improved over the last few years and shown impressive performance on synthetic data. 
However, their performance on real-world video data suffers because of the complexity of real-world degradations and misaligned video frames. 
Since obtaining a synthetic dataset consisting of low-resolution (LR) and high-resolution (HR) frames are easier than obtaining real-world LR and HR images, in this paper, we propose synthesizing real-world degradations on synthetic training datasets. The proposed synthetic real-world degradations (SRWD) include  a combination of the blur, noise, downsampling, pixel binning, and image and video compression artifacts.
We then propose using a random shuffling-based strategy to simulate these degradations on the training datasets and train a single end-to-end deep neural network (DNN) on the proposed larger variation of realistic synthesized training data. 
Our quantitative and qualitative comparative analysis shows that the proposed training strategy using diverse realistic degradations improves the performance by 7.1 \% in terms of NRQM compared to RealBasicVSR and by 3.34 \% compared to BSRGAN on the VideoLQ dataset. 
We also introduce a new dataset that contains high-resolution real-world videos that can serve as a common ground for bench-marking.

\end{abstract}
\vspace{-.7 cm}
\section{Introduction}
Video super-resolution (VSR) has extensive applications such as high-definition television, remote sensing,  surveillance systems, etc.. \cite{application_medical, application_remote, application_surrveillance}. 
Since the success of deep learning, the VSR approaches are also massively evolving, and these methods can be divided into two categories \cite{survey}:
I) Methods using synthetic datasets with pre-defined degradations for training, such as VSRNet \cite{ma_vsrnet}, FRVSR \cite{frvsr}, TecoGAN \cite{tecoGAN}, Deep-Blind-VSR \cite{pan2021deep}, BasicVSR \cite{basicvsr}, BasicVSR++, \cite{basicvsrpp} etc. 
II) Methods using real-world degradations, such as RealVSR \cite{real_2021}, and RealBasicVSR \cite{real_basic_2022}. 
The methods in category I synthesize LR images by a pre-defined degradation model (PDM) that consists of a bicubic downsampling, Gaussian blur, and noise. 
Such a degradation model is inadequate and poses a challenge for real-world VSR. 
Whereas the methods in category II, such as RealVSR \cite{real_2021}, try to overcome this issue by introducing a new real-world dataset using a multi-camera device but it lacks generalizability in diverse degradations. 
In contrast, RealBasicVSR \cite{real_basic_2022} addresses this issue by adopting a very large DNN and using a second-order degradation model \cite{real_basic_2022} that simulates degradations more than once. We observed that significantly less work \cite{real_2021, real_basic_2022} had been done toward VSR for real-world videos.

We analyzed the performance of these VSR methods on real-world datasets and have shown them in Figure \ref{cover}. The performance of the existing algorithm suffers because of the lack of diverse degradations in the training dataset, which is usually a synthetic dataset of LR-HR pairs with a pre-defined degradation or second-order degradation. 
To overcome these issues of diverse degradations
we propose to create a pool of \emph{diverse} synthetic real-world degradations (SRWD), and they are as follows:

\emph{\textbf{Blur degradation}}: Existing methods \cite{bsrgan, app_plugnplay2, iso} use isotropic and anisotropic blur kernels to simulate blur degradations whereas real-world images may have different blur patterns. To have a realistic blur kernel, we propose to extract blur kernels from real-world images using KernelGAN \cite{KernelGAN}, a self-supervised kernel extraction algorithm. 

\emph{\textbf{Noise degradation}}: In addition to Gaussian noise in the RGB space, we propose to add signal-dependent Poisson noise in the sensor RAW space to have a realistic noise distribution in the training datasets. 

\emph{\textbf{Downsampling degradation}}: We use standard downsampling operations, such as bilinear, bicubic, etc., as done in the existing methods \cite{bsrgan, app_plugnplay2, iso}.

The above degradations are applicable for both image and video modes. Below degradations are specifically for video mode, and they are as follows:

\emph{\textbf{Pixel-binning degradation}}: In the video mode, the resolution is smaller than in the image mode of the same camera. Pixel binning \cite{pixel_binning} refers to combining the electrical charges of neighboring pixels to form a superpixel, and due to this conversion, degradation is expected. We  simulate this degradation in the RAW space with the help of box filters of different sizes \cite{pixel_binning}.

\emph{\textbf{Compression artifact degradations}}: Here, we simulate degradation because of compression that arises while saving a video and we apply MPEG compression to simulate this degradation.

In this paper, we do not consider other degradations, such as motion blur, as the test datasets do not have motion blur in them, and we focus on VSR to have a fair comparison with the existing algorithms. However, the proposed SRWD can be extended to include such degradations.

\input{figures/arch_1.tex}

In summary, we propose a realistic degradation model incorporating the major source of variations in real-world data and simulate these degradations on the synthetic training data for VSR. The reason behind creating a diverse pool of SRWD is to match the characteristics of real-world videos. To simulate these degradations on the training datasets, we use the random shuffling strategy as proposed in BSRGAN \cite{bsrgan}. We then train a VSR deep neural network (DNN) on the proposed synthesized training datasets and perform extensive experiments on completely blind real-world public datasets, such as RealVSR \cite{real_2021} and VideoLQ \cite{real_basic_2022} datasets. The proposed algorithm yields objectively and subjectively superior results compared to the existing real-world VSR methods. We observe that both these public datasets contain low-resolution videos and have camera-specific degradations.  Further, to boost the VSR research community, we created a high-resolution real-world VSR dataset captured using the K$\mid$Lens \cite{klens_paper} and referred to this dataset as the K$\mid$Lens dataset.

\section{Related Work}
Since VSR algorithms trained on a pre-defined degradation model suffer on real-world videos, the RealVSR \cite{real_2021} algorithm addresses this issue by introducing a real-world VSR dataset by capturing the LR-HR video pair from a single device with a multi-camera system (Phone 11 Pro Max). Along with this dataset, the RealVSR \cite{real_2021} provides a benchmark by training different existing PDM-based-VSR algorithms on their proposed dataset. The experimental results and comparison show that these algorithms significantly improve the performance of VSR algorithms. Again, this indicates that learning using a PDM is insufficient and will suffer in real-world videos. However, the main drawback of this benchmark is that it cannot reflect a generalizable benchmark as the RealVSR dataset \cite{real_2021} consists of only degradations specifically for the iPhone camera.

Chan et al. \cite{real_basic_2022} adopt a larger DNN and propose a pre-processing stage to reduce the effect of noise and other degradations before feeding it to the DNN. They analyze how long-term propagation in the training process affects the performance of real-world videos. They proposed using a second-order degradation model \cite{realesrgan} that simulates different degradations, such as blur, noise, compression, etc., more than once. This is why the degradations are more realistic as compared to the PDM. To validate their performance on real-world datasets, they introduced a new dataset (video) with 50 LR videos obtained using web crawling. 

Unlike VSR, several image SR algorithms \cite{app_plugnplay, app_plugnplay2} focus on simulating degradations on the training datasets and achieve impressive performance. Recently, authors in BSRGAN \cite{bsrgan} proposed to apply the different degradations in random order instead of following a particular order, as shown in Figure \ref{arch_1}(b). The advantage of this strategy is that because of the random order, a different combination of degradations can be simulated, thus expanding the degradation space and training datasets. Inspired by this idea of random shuffling, we follow a similar strategy of simulating diverse degradations in random order. In addition to this random shuffling, this paper focuses on expanding the degradations space by extracting blur kernels from real-world images, signal-dependent sensor noise, pixel-binning, and video compression artifacts. With a diverse pool of SRWD, we then use a random-shuffling-based strategy to create the LR/HR training data. 
Our experimental results and analysis sis shows that the random-shuffling strategy-based SRWD can model real-world degradations better than the existing algorithms.

\vspace{-.2 cm}
\section{Proposed SRWD-VSR Algorithm}
In the following subsections, we describe our realistic degradation model (Section~\ref{sec:degradation_model}) and VSR architecture (Section~\ref{sec:VSR_model}) for our SRWD-VSR algorithm.

\subsection{Realistic Degradation Model}
\label{sec:degradation_model}
The super-resolution process aims at recovering a high-resolution (HR) image from the given low-resolution (LR) image. In general,  the LR image ($I^{LR}$) can be mathematically modeled as,

\begin{equation}
I^{LR} = \Delta(I^{HR};\delta)
\end{equation}

\input{figures/arch_2.tex}
\input{figures/kernels.tex}

where $\Delta$ denotes the mapping function of degradation, $I^{HR}$ is the corresponding HR image, and $\delta$ represents the various parameters of the degradation process (e.g., blur, downsampling, noise, etc.). Although in real-world images/videos, $\Delta$ and $\delta$ are unknown, researchers try to imitate the $\delta$ parameters such that the degradation $\Delta$ is as close as possible to the real-world images. As discussed earlier, also, towards real-world VSR, two approaches have been proposed in the literature. RealVSR \cite{real_2021} tries to get a real $I^{HR}$ and $I^{LR}$ pair by using a multi-camera device, whereas RealBasicVSR \cite{real_basic_2022} trains their model on synthetically created $I^{HR}$ and $I^{LR}$ pairs using a second order degradation model. However, the degradation space is still limited and has the following issues:

\emph{1)} The blur kernels used in all the existing degradation models are standard isotropic or anisotropic kernels, whereas real-world images may have different blur patterns \cite{KernelGAN}.
\emph{2)} The real sensor noise is very different from the Gaussian Noise, which is used as a noise factor in these degradation models \cite{cycleISP}.
\emph{3)} Also, not performing any blur operation on RAW space to consider the effect of pixel-binning.

Considering all these points in context with VSR, we propose to expand the degradation space, and we list them below. An overview of the proposed realistic degradation model is shown in Figure \ref{arch_2}.
\vspace{-.3cm}
\paragraph{Blur Kernel Pool.} Most degradation models rely on fixed blur kernels, such as Gaussian kernel filters \cite{realesrgan}, to get a synthesized $I^{LR}$ image from a given $I^{HR}$ image. In contrast, real-world LR images do not comply with this assumption. To match the blurring characteristics in real-world images, we propose to extract blur kernels from the real-world images using KernelGAN \cite{KernelGAN}. The KernelGAN \cite{KernelGAN} is a dataset invariant, fully unsupervised, and single input image method which can predict the blur kernel in the wild. KernelGAN consists of a generator (G) and a discriminator (D) to downscale. This fully-convolutional G and D work together to predict the SR kernel of the given image. At a patch level, G learns to downscale the given image so that, for D, it is indistinguishable from the image. 

In this paper, we create a kernel pool of approximately 5000 kernels from a dataset of 5000 images collected from different sources for generality and we refer to them as ($B_{real}$). Figure \ref{kernels} shows some examples of these kernels. 
\vspace{-.4 cm}
\paragraph{Sensor Noise Addition.}
Simple noise models like additive white Gaussian noise are not enough for a realistic degradation model as the real-world color images are complicated. Collecting real noisy and clean images is tedious and unsuitable for large-scale learning. We propose injecting realistic noise in the raw sensor space by first converting the image from sRGB to raw space. 

Most VSR methods use Gaussian noise and/or a step-by-step technique of inverting camera image signal processing (ISP) to convert sRGB into raw data \cite{bsrgan}. The drawback of such an approach is that it needs prior information on the targeting camera device. It is also challenging to reverse several camera imaging pipeline engineering operations. In the proposed degradation model, we use CycleISP \cite{cycleISP}, an image noise synthesizer that uses a Poisson-Gaussian noise model\cite{cycle_ref} and can generate realistic synthetic clean/noisy paired data in raw and sRGB spaces. We name the injected noise as $N_{real}$.

\vspace{-.2 cm}
\paragraph{Traditional degradations.} Apart from the above-mentioned realistic blur and noise models, adding traditional down-sampling, noise, and blurring can be used for augmentation and make the degradation model more comprehensive \cite{bsrgan}. Therefore, the proposed degradation model also uses the following traditional degradations:

\emph{\textbf{Gaussian Blur:}} In the proposed degradation model, we perform two Gaussian blur operations: $B_{iso}$ with isotropic blur kernels and
$B_{aniso}$ with anisotropic blur kernels.

\emph{\textbf{Down-sampling:}} There are three types of downsampling ways used in the proposed degradation model, i.e., nearest neighbor interpolation ($D^s_{nearest}$), bilinear interpolation ($D^s_{bilinear}$), bicubic interpolation ($D^s_{bilinear}$), down-up sampling ($D^s_{down-up}$), where $s$ is the scaling factor.

\emph{\textbf{Gaussian and JPEG compression noise:}} We adopted the three-dimensional zero-mean Gaussian noise model ($N_G$) and JPEG compression noise ($N_{JPEG}$) with quality factors ranging [30, 95] in the proposed degradation model.

\paragraph{Pixel Binning.} In general, blur operations are conducted on sRGB instead of sensor RAW space. Furthermore, in the video mode, the resolution is smaller than in the image mode for the same camera. This decrease in resolution for the same sensor size is achieved by a process commonly referred to as pixel binning \cite{pixel_binning}. Combining the electrical charges of neighboring pixels to form a super-pixel is termed pixel binning \cite{pixel_binning}. The main benefit of this technique is that the combined charges would overcome the real noise at the expense of spatial resolution. To simulate the effect of pixel binning, we use box filters of different dimensions for this purpose. The sizes for box filters are randomly chosen to range from 3$\times$3 to 15$\times$15. 

\paragraph{Video Compression.} Once all the degradations are simulated, we apply video compression to LR frames to generate artifacts due to compression. Unlike image degradations, video compression implicitly considers the inter-dependencies between video frames, providing us with temporally and spatially varying degradations \cite{real_basic_2022}. Therefore, in each iteration, we randomly selected one of the following codecs: ``libx264", ``h264", ``mpeg4", and a bitrate ranging between $[10^4, 10^5]$.

\input{figures/frvsr_arch.tex}
\subsubsection{Random Shuffling Strategy} In this paper, we simulate the degradations mentioned above in a random-shuffling strategy as proposed in BSRGAN \cite{bsrgan} that helps to expand the degradation space and is depicted in Figure \ref{arch_2}. As shown in Figure \ref{arch_2}, the proposed SRWD pool contains \emph{\textbf{1)}} Blur Kernels consisting of real-world kernels $B_{real}$ as well as isotropic ($B_{iso}$) and anisotropic ($B_{aniso}$) kernels, \emph{\textbf{2)}} Downsampling operations ($D^{s}$), \emph{\textbf{3)}} Noise ($N_{real}$, $N_G$, $N_{JPEG}$), \emph{\textbf{4)}} Pixel binning and \emph{\textbf{5)}} video compression. 

For a given clean HR video frame ($I^{HR}$) with no degradations, the proposed algorithm shuffles these eight degradations in a random order, followed by video compression, and applies each of them to generate a realistic synthesized $I^{LR}$. More specifically, these 8 degradations are : $B_{real}$, $B_{iso}$, $B_{aniso}$, $N_{real}$, $N_G$, $N_{JPEG}$, $D^{s}$ and pixel binning. 

\subsection{Video Super Resolution Model} \label{sec:VSR_model}

The proposed SRWD model is utilized to simulate a training dataset for a blind VSR and can be used to enhance the performance of any video super-resolution algorithm. For experiments and analysis, in this paper, we adopt the widely-used FRVSR model \cite{frvsr} as a backbone architecture, with some changes to account for varied, realistic degradation and shuffling strategies. Since the degradations are complex and complicated, it is very vital to have the right sub-module in the architecture and this is why we propose to optimize each of the components in order to have a better super-resolution. 
In the following, we describe the architecture used for training shown in Figure~\ref{frvsr_architecture}, and the important blocks of the architecture are described below.

\textbf{Optical Flow.} To consider severe degradations, we replace the optical flow estimation module  with FlowNet2 \cite{flownet2}, which is robust to degradations, such as noise and blur.

\textbf{Bidirectional Flow.} The bidirectional propagation has proven to help learn the VSR better than unidirectional\cite{basicvsr}. Hence, the unidirectional propagation of FRVSR is replaced by bidirectional propagation, i.e., each frame receives accumulated information from both directions (forward and backward). 

\textbf{Feature Warping.} The image space warping is also replaced with the feature space warping. The reason behind this is that in the image space, the warped images inevitably suffer from blurriness and incorrectness due to the inaccuracy of optical flow estimation \cite{basicvsr}.

\textbf{Super resolution.} The upsampling module is replaced by RRDBNet \cite{esrgan,cited_during_rebuttal} to enhance the performance of FRVSR architecture. The first convolutional block of RRDBNet has also been changed to take 128 channel frames as input instead of 3 channel frames because of the introduction of the feature warping module. We replace the VGG-based discriminator with a U-Net-like discriminator with skip connections \cite{realesrgan}. The normalized spectral regularization stabilizes the training dynamics, and this architecture can also provide per-pixel feedback to the generator. 

We then train the above VSR architecture on the simulated training datasets using loss functions in \cite{realesrgan}, which is a weighted combination of content loss, perceptual loss, and vanilla adversarial loss.

\input{figures/k_lens.tex}

\section{Proposed K$|$Lens datasets}
RealVSR \cite{real_2021}, and VideoLQ \cite{real_basic_2022} are two publicly available blind VSR datasets in the literature. 
The RealVSR testing dataset consists of 50 videos, 50 frames each video, $512\times1024$ dimension each frame. The VideoLQ dataset also contains 50 videos downloaded from various video-hosting websites. The videos are approximately 100 frames each and have $640\times480$ resolution for each frame. The videos of the RealVSR dataset are captured using an iPhone camera and hence contain only phone camera-specific degradations, and both these datasets have low-resolution videos. 

To further boost the VSR community and to have a high-resolution video dataset, in this paper, we propose introducing a new dataset that contains real-world videos captured using K$|$Lens \cite{klens_paper} camera.
The unique optical lens developed by K$|$Lens \cite{klens_paper} enables any camera with exchangeable lenses to capture multiple perspectives of a scene with a single exposure as regular color images on the camera sensor. It is a new optical lens, and the images or videos shot by K$|$Lens contain 9 perspectives of a scene, and most importantly, 
we observe that the number of degradations, such as noise and blur, also vary among these nine perspectives. Thus, shooting one scene results in 9 videos and each of these 9 videos differs in perspectives and amount of degradations. This is one of the reasons why we chose to create a video-SR dataset with K$|$Lens to have diverse levels of degradations in the datasets. The proposed K$|$Lens dataset contains 56 videos with 100 frames each with resolution $1600\times1080$ resolution videos. The dataset consists of indoor and outdoor scenes with varied motions, degradations, and lighting conditions. 
We have shown a few example scene shots with K$|$Lens in Fig. \ref{k_lens}.  
Compared to existing datasets, the proposed K$|$Lens dataset has high-resolution videos with diverse degradation levels.
\input{tables/quant_self}

\section{Experiments and Analysis}
In this section, we extensively perform experiments on public datasets, such as VideoLQ \cite{real_basic_2022} and RealVSR \cite{real_2021}, and also show comparisons on the proposed K$|$Lens datasets. We show quantitative and quality comparisons in
our experiments for analysis and discussion.

\input{figures/qual_self.tex}
\subsection{Training}
\textbf{Training datasets} Our training dataset combines the Vimeo \cite{vimeo} and REDS \cite{reds} datasets. Vimeo consists of 277 video scenes, each having 120 frames, and REDS consists of 240 HD scenes, each having 100 frames. We extract patches of size 128 $\times$ 128 for scale $2$ ($2$ times SR) and 256 $\times$ 256 for scale $4$ from these datasets to generate HR videos. We consider these HR datasets and apply the proposed SRDW model to produce the corresponding LR videos. We use 10 consecutive frames of a video and use a batch size of 4, so a total of 40 images are processed in a batch. For degradation simulations with a batch size of 4, each having 10 frames, the first 10 frames will be blurred with the same kernel, same downsampling operations, and have the same noise level setting in addition to pixel-binning and video compression, and the second group of 10 frames may have different blur kernels, noise, etc. from the first group. This is done to avoid a case where subsequent frames with two extreme levels of blur, i.e., no blur and heavy blur, which is not a realistic scenario. Data augmentation consisting of random horizontal and vertical flips is also applied for better generalization.
\input{tables/quant_self_degra}
\textbf{Training details}  We implement the network in PyTorch and leverage the built-in Adam optimizer \cite{adam} with $\beta_1$ = 0.9 and $\beta_2$ = 0.99 and an initial learning rate of $5 \times 10^{-5}$. To speed up the training, we freeze the weights of Flownet2 \cite{flownet2} as it is already pre-trained and initialize the RRDB network with the model \cite{bsrgan} and train the network for $75,000$ iterations. Similar to this \cite{realesrgan}, all degradation operations are implemented with CUDA acceleration to synthesize training pairs on the fly. A training queue is maintained to increase the degradation diversity in the batch. The queue size is set to $180$.

\input{figures/qual_sota}

\input{tables/quant_sota}

\vspace{.4cm}
\subsection{Results and comparison}
 Here, we adopted three real-world datasets: RealVSR \cite{real_2021}, VideoLQ and \cite{real_basic_2022}, proposed K$|$Lens \cite{klens_paper} for our experimental analysis. All these three datasets are real-world datasets without ground truths. Therefore, similar to RealBasicVSR \cite{real_basic_2022}, no-reference quality assessment metrics are adopted as the comparison method in this paper. We used NRQM \cite{nrqm} and BRISQUE \cite{brisque} metrics for this purpose. The higher the value of NRQM better is the performance of the VSR algorithm, while the lower the value of BRISQUE better is the VSR algorithm.
 
To demonstrate the effect of the proposed SRWD model and modified architecture, we train the proposed algorithm with different settings:

\textbf{SRWD-VSRv1:} This version is trained with the proposed SRWD model but without architecture modification and no shuffling. 

\textbf{SRWD-VSRv2:} This version is trained with the proposed SRWD model, with proposed architecture modification but with no shuffling.

\textbf{SRWD-VSR:} In this version, shuffling is performed on the proposed SRWD model with the architecture modifications.

Table \ref{qual_self_table} summarizes the performance of all these versions on three datasets. It can be observed from the table that the proposed SRWD-VSR is better in all the datasets. We have also shown the visual comparisons between these versions to demonstrate the qualitative analysis in Figure \ref{qual_self}. It can be observed that the results of SRWD-VSR are more visibly plausible than the compared versions. For instance, only SRWD-VSR could generate patches with negligible artifacts (see Figure \ref{qual_self} (e), (j), (o)). The above experiments give an analysis of the contribution of each of the SRWD models, shuffling-based strategy, and architecture modifications on the proposed SRWD-VSR. This analysis also quantifies the performance gains coming from the architecture modifications as compared to the ones coming from the proposed degradation pipeline. Next, we study the effect of the different degradations in the SRWD model and train the proposed algorithm with different settings as described below:

\textbf{SRWD-VSR-KG:} In this version, we analyze the influence of KernelGAN over isotropic and anisotropic kernels. Instead of generating kernels from KernelGAN, we generate 5000 kernels of isotropic and anisotropic kernels by tuning the variance parameters. In this version, all the proposed degradations are taken into account except 5000 real-world kernels are replaced with 5000 isotropic and anisotropic kernels. Here, we use the proposed architecture with shuffling-based degradations simulation.

\textbf{SRWD-VSR-SN:} In this version, we analyze the influence of sensor noise. Here, all the degradations are taken into account except sensor noise and we use the proposed architecture modifications and shuffling.

\textbf{SRWD-VSR-PB:} In this version, we discuss the influence of pixel-binning and video compression operations. Here, again all the degradations are taken into account except pixel-binning and video compression degradations and we use proposed architecture modifications and shuffling. 

Table \ref{qual_self_table_deg} summarizes the performance of all these versions on three datasets. Here, SRWD-VSR is the case where all the degradations are taken into account with shuffling and architecture modification. This experiment quantifies the performance gains coming from the different degradations and the removal of any one of the degradations will reduce the performance of the proposed SRWD-VSR algorithm.

\textbf{Comparison to State-of-the-Art (SOTA)} We compare our SRWD-VSR method with several state-of-the-art methods, including RealBasicVSR\cite{real_basic_2022}, RealVSR\cite{real_2021}, TecoGAN\cite{tecoGAN}, BSRGAN\cite{bsrgan}, Real-ESRGAN\cite{realesrgan}.
Since TecoGAN \cite{tecoGAN} uses the FRVSR \cite{frvsr} generator with an additional discriminator and trains its algorithm with sophisticated loss functions, thus we include TecoGAN, not FRVSR, in the experimental sections and analysis.

We compare these existing state-of-the-art methods subjectively and objectively on scenes from all three datasets, and the results are shown in Table \ref{qual_sota_table} and \ref{qual_sota_fig}. The proposed algorithm generates much more detail in fine regions, improving visual quality, as seen in Figure \ref{qual_sota_fig}. Regarding comparison using no-reference performance parameters, i.e., NRQM and BRISQUE, the proposed SRWD-VSR performs better than all the five super-resolution methods for all three tested datasets and is shown in Table \ref{qual_sota_table}. 
Thus, we can conclude that the proposed algorithm is better in all the datasets.

\vspace{-.2cm}

\section{Conclusion}
LR-HR pairs' formation for VSR algorithms' training plays an important role, especially for real-world videos. In this paper, we have investigated a realistic degradation model for video super-resolution methods by introducing a synthetic dataset of real-world degradations and using a random shuffling strategy to train a state-of-the-art VSR DNN architecture. We observed that none of the existing degradation models could imitate the complicated real-world degradations in its LR-HR pairs for training their VSR models to the same extent as ours. The experimental analysis conducted in this paper suggests that the proposed SRWD-VSR performs qualitatively and quantitatively better than all the existing VSR models in the literature.

{\small
\bibliographystyle{ieee_fullname}
\bibliography{main}
}

\clearpage
\newpage
\pagebreak

\begin{center}
  \huge{\bfseries --- Supplementary ---}\\
\end{center}

\section{ Blur Kernel Pool ($B_{real}$):}
Recall that the proposed algorithm creates a blur kernel pool from real-world images. We propose to use KernelGAN \cite{KernelGAN} for this purpose and create a kernel pool of approximately 5000 kernels from a dataset of 5000 images. These 5000 images are collected from DF2K \cite{esrgan}, K$|$Lens datasets \cite{klens, klens_paper}. DF2K is an image dataset, whereas K$|$Lens datasets consist of videos. 

Please note that in the training of our algorithm, for creating an LR-HR pair, we randomly selected one blur kernel out of these 5000 blur kernels, in addition to isotropic and anisotropic blur kernels  for simulation.

\section{Proposed K$|$Lens Dataset:}
The unique optical lens developed by K$|$Lens \cite{klens, klens_paper} enables any camera with exchangeable lenses to capture multiple perspectives of a scene with a single exposure as regular color images on the camera sensor. More specifically, it captures nine different perspectives of the same scene and also it can record videos. Please refer to \cite{klens, klens_paper} for more details.


\section{ Quality Assessment Metrics:} We have used two no-reference quality assessment metrics for quantitative comparison in the paper, i.e., NRQM \cite{nrqm}, and BRISQUE \cite{brisque}. BRISQUE is a widely used metrics originally proposed for the quality evaluation of natural images. These methods rely upon natural scene statistical (NSS) features extracted from local image patches to calculate the quality of the distorted image. The BRISQUE is trained on features obtained from natural and distorted images and human judgments.

NRQM\cite{nrqm} is specifically designed to predict the quality scores of super-resolved images. They proposed to use three types of low-level statistical features in both spatial and frequency domains. These features are learned using a two-stage regression model to predict the quality scores of the super-resolved image without referring to ground-truth images. Extensive experimental analysis has been done in the \cite{nrqm} to compare NRQM with the existing no-reference IQA metrics, including BRISQUE metric. In terms of Spearman Rank Correlation Coefficient (SRCC), \cite{srcc}, the NRQM metric is better than all the compared metrics, as suggested in the \cite{nrqm}. Recall that we extensively compared different VSR algorithms in Tables 1, 2 and 3 of the manuscript, and we can observe that the proposed SRWD-VSR outperforms all the compared models in terms of NRQM.

\section{Pre-trained weights of RealBasicVSR \cite{real_basic_2022}}
Recall that in the simulation results and experiments (Table 3 of the main paper), we have not shown the results \cite{real_basic_2022} for $2$ $\times$ scaling as we do not have the pre-trained weights. This is why visual results in Figure 8 of the main paper are only based on $4$ $\times$ scaling.

\end{document}

%% file: figures/cover.tex
\vspace{-2cm}
\begin{strip}
\begin{minipage}{.195\textwidth}
  \centering
  \includegraphics[width=0.99\linewidth, height=45mm]{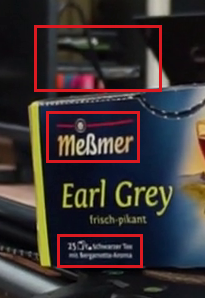}  
\captionof*{figure}{LR Input}
\end{minipage}
\begin{minipage}{.195\textwidth}
\includegraphics[width=\linewidth]{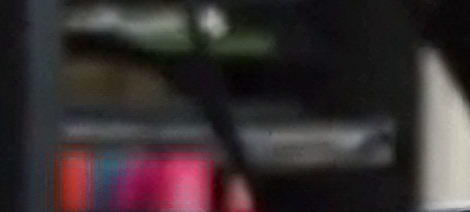}  \\
  \includegraphics[width=\linewidth]{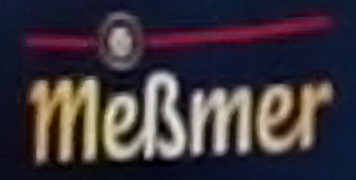}  \\
     \includegraphics[width=\linewidth]{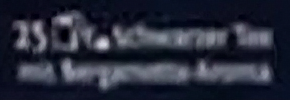} 
    \captionof*{figure}{TecoGAN\cite{tecoGAN}}
\end{minipage}
\begin{minipage}{.195\textwidth}
     \includegraphics[width=\linewidth]{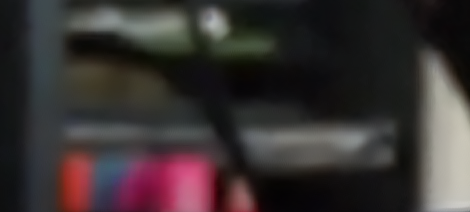}  \\
  \includegraphics[width=\linewidth]{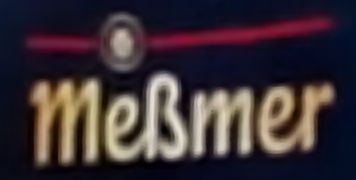}  \\
     \includegraphics[width=\linewidth]{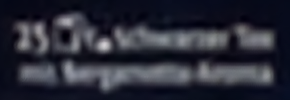} 
    \captionof*{figure}{RealVSR\cite{real_2021}}
\end{minipage}
\begin{minipage}{.195\textwidth}
      \includegraphics[width=\linewidth]{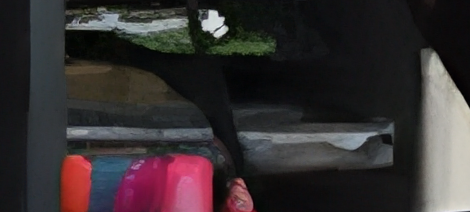}  \\
  \includegraphics[width=\linewidth]
  {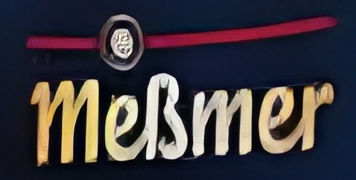}  \\
     \includegraphics[width=\linewidth]{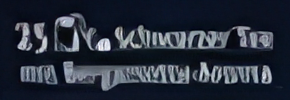} 
    \captionof*{figure}{RealBasicVSR\cite{real_basic_2022}}
\end{minipage}
\begin{minipage}{.195\textwidth}
  \includegraphics[width=\linewidth]{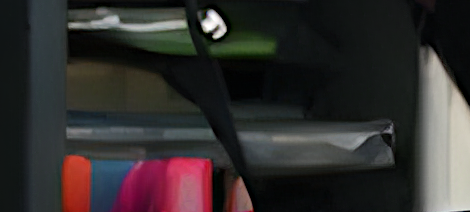}  \\
  \includegraphics[width=\linewidth]{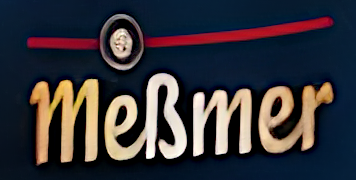}  \\
     \includegraphics[width=\linewidth]{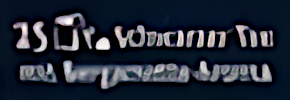} 
     \captionof*{figure}{SRWD-VSR(ours)}
\end{minipage}
\captionof{figure}{\textbf{Results on real-world video.} This figure compares the proposed SRWD-VSR algorithm with the existing state-of-the-art VSR methods ($\times$ 4 SR). This work shows how varied random degradations can contribute to learning an effective VSR model, especially for real-world video artifacts. \textbf{(Zoom-in for best view)}}
\label{cover}
\end{strip}

%% file: figures/arch_1.tex
\begin{figure}[t]
\setcounter{figure}{1}
\begin{center}
\includegraphics[width=86 mm]{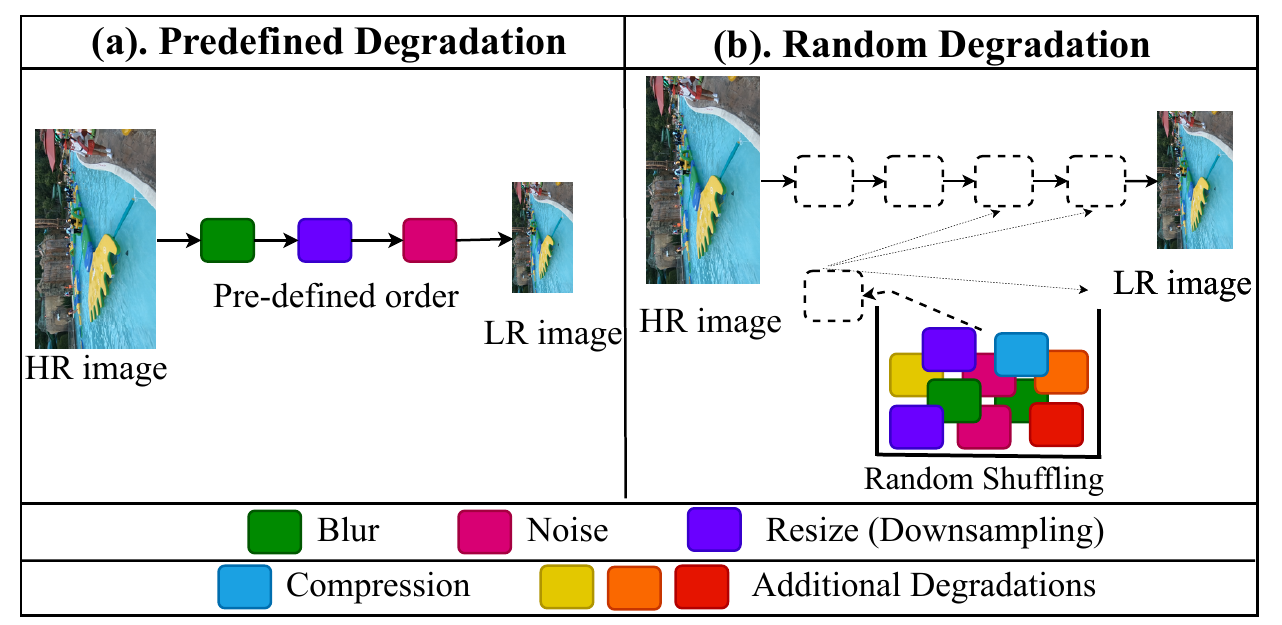}
\end{center}
   \caption{\textbf{Synthetic LR-HR pair creation.} 
  Our approach is based on using random degradation techniques, as shown in (b). 
   }
   \label{arch_1}
\end{figure}

%% file: figures/arch_2.tex
\begin{figure}[t]
\begin{center}
\includegraphics[width=81mm]{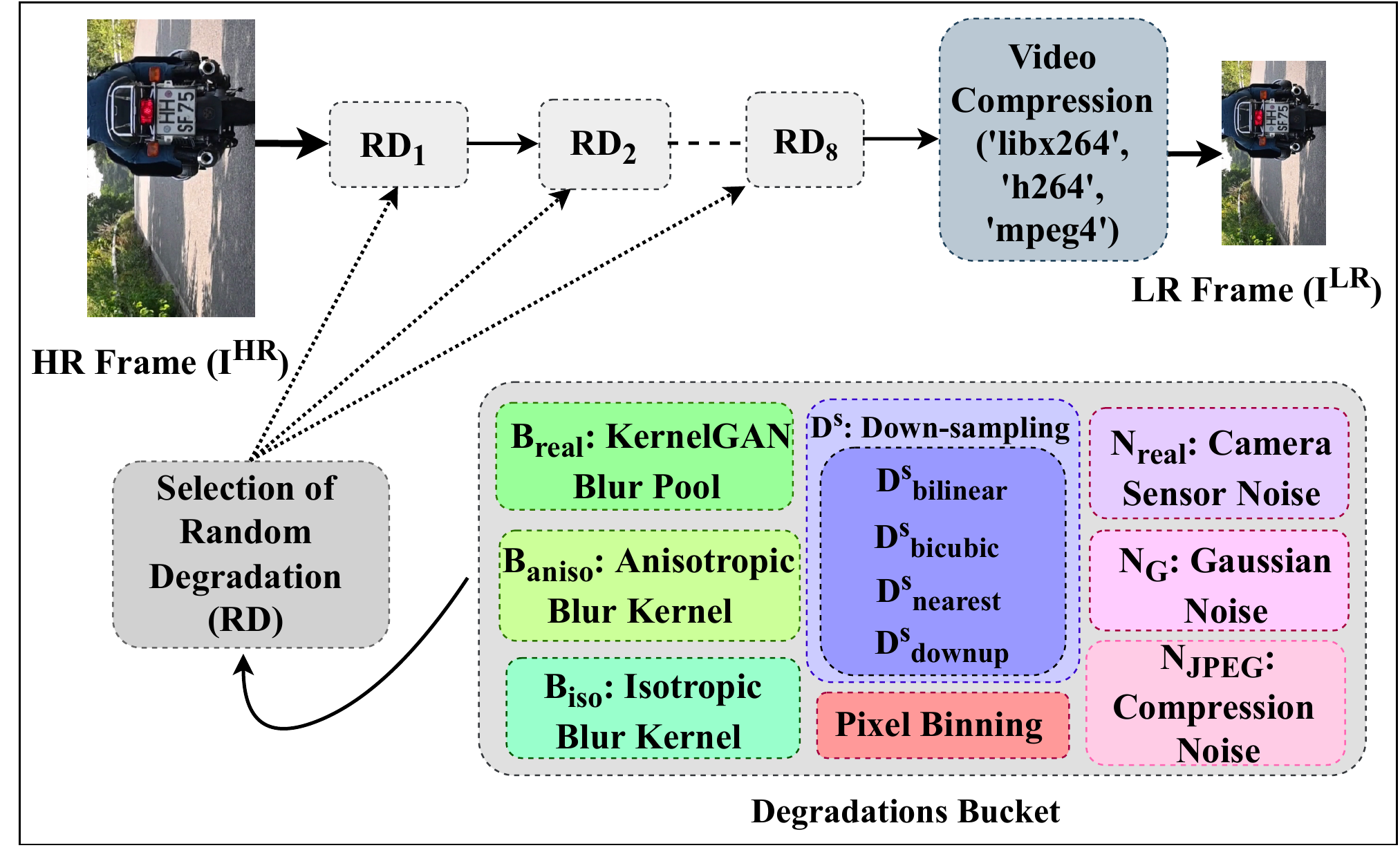}
\end{center}
   \caption{\textbf{Proposed SRWD model.} The degradation bucket contains all eight types of degradations from which the random selection is done, which are then followed by video compression.}
\label{arch_2}
\end{figure}

%% file: figures/kernels.tex
\begin{figure}[th]
\centering
\includegraphics[width=\linewidth]{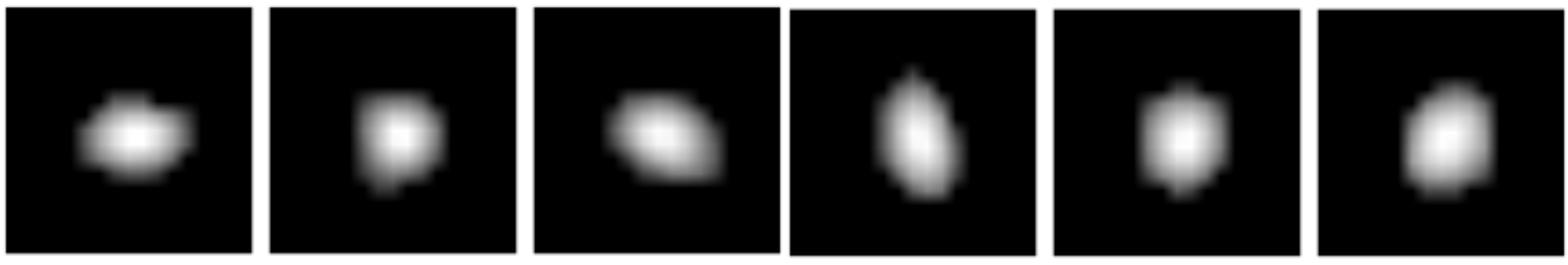}
   \caption{\textbf{Kernel Pool.} Some examples from the proposed kernel pool. These kernels are extracted from 6 real-world images using the KernelGAN method \cite{KernelGAN}.}
   \label{kernels}
\end{figure}

%% file: figures/frvsr_arch.tex
\begin{figure}[t]
\begin{center}
\includegraphics[width=81mm]{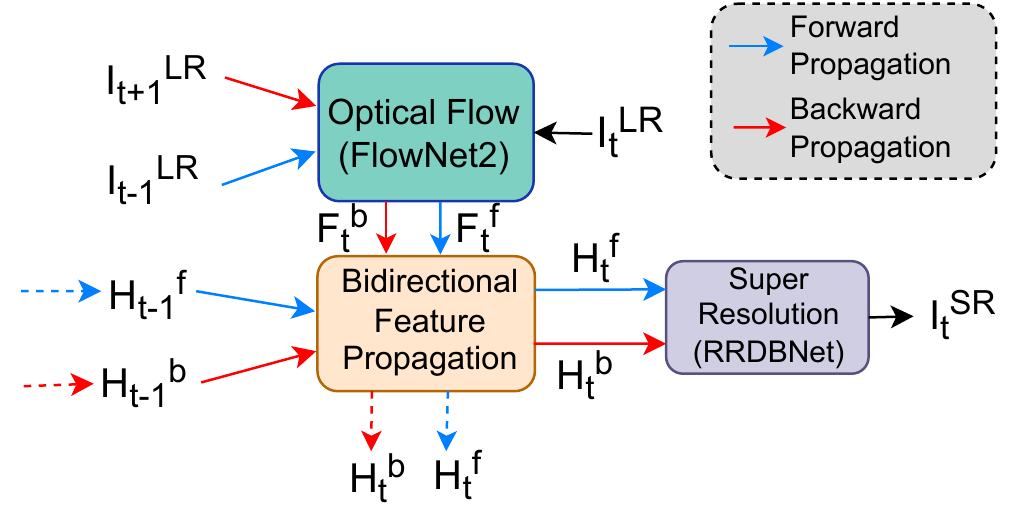}
\end{center}
   \caption{\textbf{Proposed VSR Architecture.} On a high level, we calculate the optical flow of the current frame $I^{LR}_t$ with respect to the previous $I^{LR}_{t-1}$ and the next $I^{LR}_{t+1}$ frames. We warp the features $H^{f}_{t-1}$ and $H^{b}_{t-1}$ based on the flow calculated ($F^f_t$ and $F^b_t$) in the Bidirectional Feature Propagation Module and pass the outputs $H^{f}_{t}$ and $H^{b}_t$ to the super-resolution network. The features $H^{f}_{t}$ and $H^{b}_t$ are channel-wise concatenated and fed to the RRDBNet for upsampling.}
\label{frvsr_architecture}
\end{figure}

%% file: figures/k_lens.tex
\begin{figure}[th]
\begin{subfigure}{.235\textwidth}
  \centering
  \includegraphics[width=\linewidth]{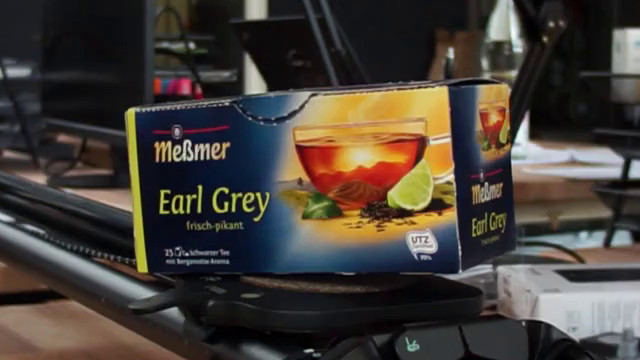}  \\
  \includegraphics[width=\linewidth]{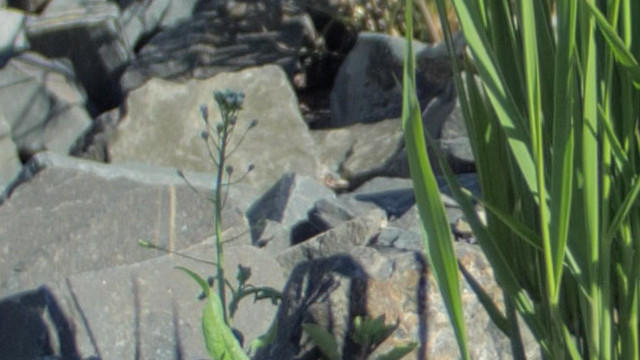}  \\
  \includegraphics[width=\linewidth]{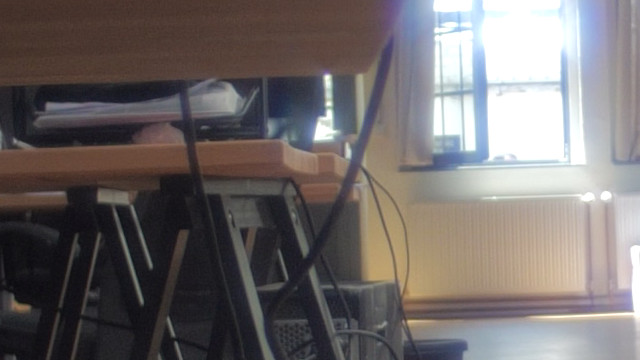} 
\end{subfigure}
\begin{subfigure}{.235\textwidth}
  \centering
  \includegraphics[width=\linewidth]{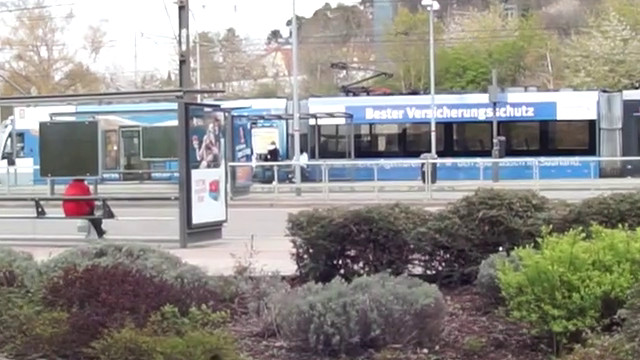}  \\
    \includegraphics[width=\linewidth]{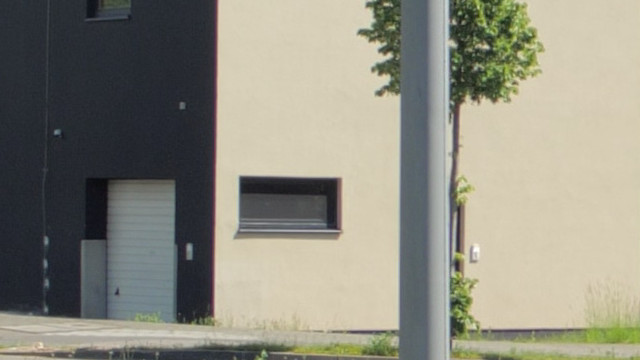}  \\
    \includegraphics[width=\linewidth]{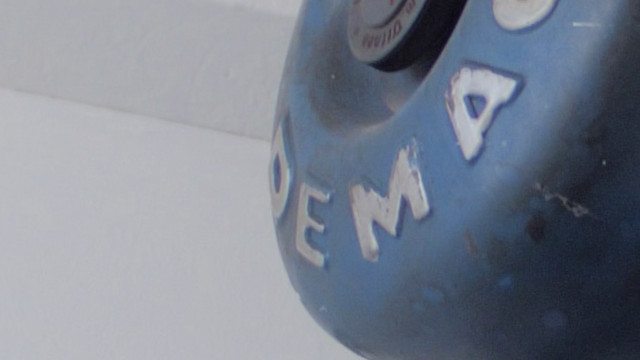}  
   \end{subfigure} 

\caption{\textbf{Example frames from the proposed K$|$Lens dataset.} These videos are taken from K$\mid$Lens camera with varied scenes (indoor/outdoor) and object motions. }
\label{k_lens}
\end{figure}

%
%


%% file: tables/quant_self.tex
\begin{table}[t]
\centering
\renewcommand{\arraystretch}{1.1}
\setlength{\tabcolsep}{2pt}
\caption{\textbf{Step-wise performance of our VSR algorithm.} Symbols `$\uparrow$' indicates a higher value is better while `$\downarrow$' indicates a lower value
is better. Overall, SRWD-VSR performs consistently better. (Note: These results are for $\times$ 2 super-resolution)}
\begin{tabular}{|C{.3cm}|C{2.3cm}||C{1.5cm}|C{1.7cm}|}
\hline
\textbf{} & \textbf{Algorithm} &  \textbf{NRQM$\uparrow$} & \textbf{BRISQUE$\downarrow$}  \\
\hline \hline
\multirow{4}{*}{\rotatebox[origin=c]{90}{K$\mid$Lens}} & SRWD-VSR  & \textbf{7.1013} & \textbf{26.9868}  \\

& SRWD-VSRv2  & 6.8011 & 27.3214 \\

& SRWD-VSRv1 & 6.5914  & 27.9245  \\


& Bicubic  & 3.3132  & 48.9655  \\ \hline

\hline
\multirow{4}{*}{\rotatebox[origin=c]{90}{RealVSR}} & SRWD-VSR & \textbf{6.7507} & \textbf{29.1829}  \\

& SRWD-VSRv2  & 6.4121  & 32.8701  \\

& SRWD-VSRv1 & 6.2120 & 34.9144  \\


& Bicubic  & 3.3715  & 55.8189  \\ \hline

\hline
\multirow{4}{*}{\rotatebox[origin=c]{90}{VideoLQ}} & SRWD-VSR & \textbf{6.9054}  &  \textbf{29.4897} \\

& SRWD-VSRv2 & 6.8012  & 30.3145  \\

& SRWD-VSRv1 & 6.7111  & 31.9148  \\

& Bicubic  & 4.0072  & 45.0282  \\ \hline
\end{tabular}
\label{qual_self_table}
\end{table}

%% file: figures/qual_self.tex
\begin{figure*}[th]
\renewcommand{\arraystretch}{1.02}
\begin{subfigure}{.3\textwidth}
  \centering
  \includegraphics[width=0.8\linewidth, height=37mm]{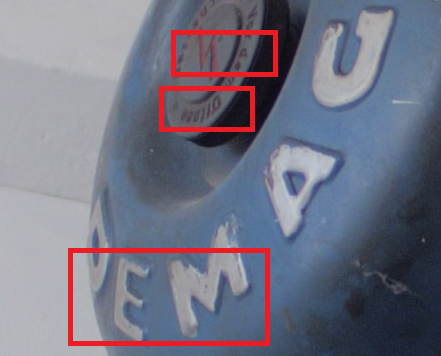}  
  \caption{K$|$Lens Dataset}
\end{subfigure}
\begin{subfigure}{.17\textwidth}
  \centering
      \includegraphics[width=\linewidth]{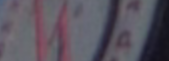} \\
  \includegraphics[width=\linewidth]{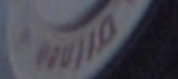}  \\
    \includegraphics[width=\linewidth]{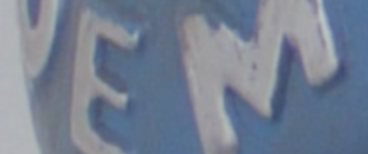}  
  \caption{Bicubic}
\end{subfigure}
\begin{subfigure}{.17\textwidth}
  \centering
      \includegraphics[width=\linewidth]{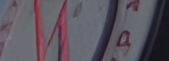} \\
  \includegraphics[width=\linewidth]{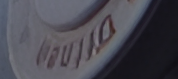}  \\
    \includegraphics[width=\linewidth]{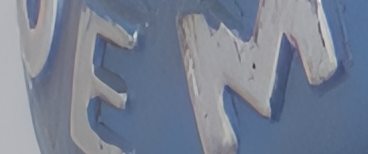}  
  \caption{SRWD-VSRv1}
\end{subfigure}
\begin{subfigure}{.17\textwidth}
  \centering
      \includegraphics[width=\linewidth]{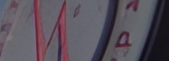} \
  \includegraphics[width=\linewidth]{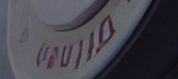}  \\
    \includegraphics[width=\linewidth]{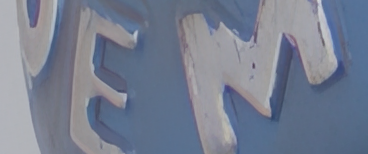}  
  \caption{SRWD-VSRv2}
\end{subfigure}
\begin{subfigure}{.17\textwidth}
  \centering
      \includegraphics[width=\linewidth]{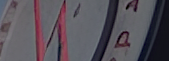} \\
  \includegraphics[width=\linewidth]{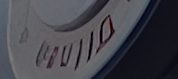}  \\
    \includegraphics[width=\linewidth]{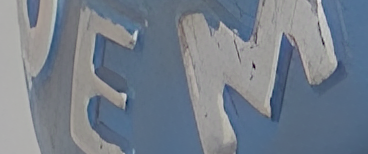} 
  \caption{SRWD-VSR}
\end{subfigure}
\newline
\begin{subfigure}{.3\textwidth}
  \centering
  \includegraphics[width=.8\linewidth, height=37mm]{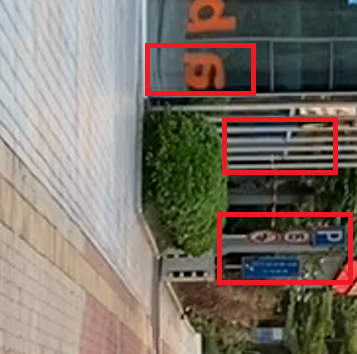}  \caption{RealVSR dataset}
\end{subfigure}
\begin{subfigure}{.17\textwidth}
  \centering
  \includegraphics[width=\linewidth]{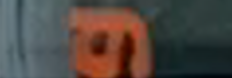}  \\
      \includegraphics[width=\linewidth]{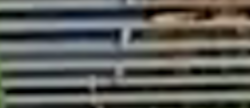} \\
    \includegraphics[width=\linewidth]{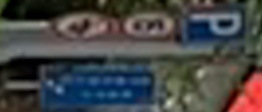}  
  \caption{Bicubic}
\end{subfigure}
\begin{subfigure}{.17\textwidth}
  \centering
  \includegraphics[width=\linewidth]{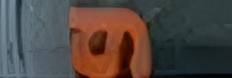}  \\
       \includegraphics[width=\linewidth]{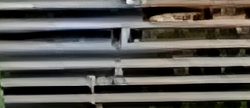} \\
    \includegraphics[width=\linewidth]{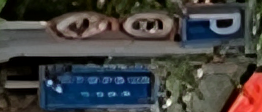}  
  \caption{SRWD-VSRv1}
\end{subfigure}
\begin{subfigure}{.17\textwidth}
  \centering
  \includegraphics[width=\linewidth]{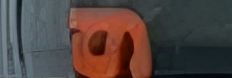}  \\
      \includegraphics[width=\linewidth]{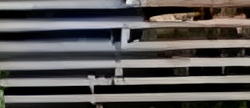} \\
    \includegraphics[width=\linewidth]{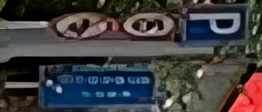}  
  \caption{SRWD-VSRv2}
\end{subfigure}
\begin{subfigure}{.17\textwidth}
  \centering
  \includegraphics[width=\linewidth]{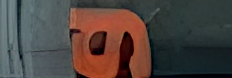}  \\
       \includegraphics[width=\linewidth]{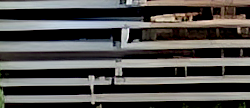} \\
    \includegraphics[width=\linewidth]{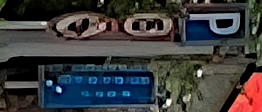} 
  \caption{SRWD-VSR}
\end{subfigure}
\newline
\begin{subfigure}{.3\textwidth}
  \centering
  \includegraphics[width=.8\linewidth, height=37mm]{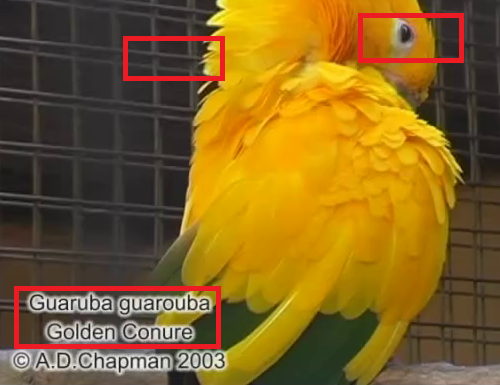}
  \caption{VideoLQ dataset}
\end{subfigure}
\begin{subfigure}{.17\textwidth}
  \centering
    \includegraphics[width=\linewidth]{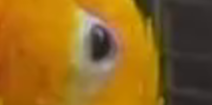}  \\
    \includegraphics[width=\linewidth]{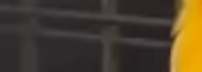} \\
      \includegraphics[width=\linewidth]{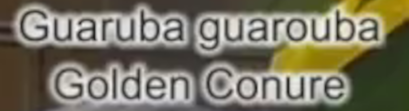}  
  \caption{Bicubic}
\end{subfigure}
\begin{subfigure}{.17\textwidth}
  \centering
    \includegraphics[width=\linewidth]{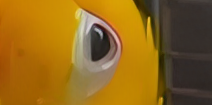}  \\
     \includegraphics[width=\linewidth]{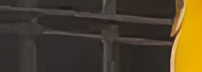} \\
       \includegraphics[width=\linewidth]{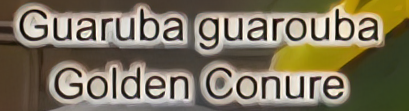}  
  \caption{SRWD-VSRv1}
\end{subfigure}
\begin{subfigure}{.17\textwidth}
  \centering
    \includegraphics[width=\linewidth]{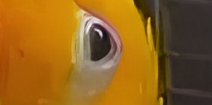}  \\
     \includegraphics[width=\linewidth]{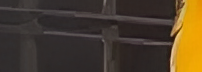} \\
       \includegraphics[width=\linewidth]{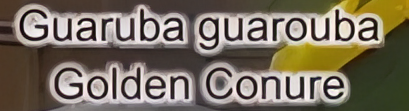}  \\
  \caption{SRWD-VSRv2}
\end{subfigure}
\begin{subfigure}{.17\textwidth}
  \centering
    \includegraphics[width=\linewidth]{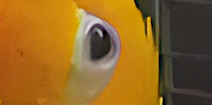} \\
     \includegraphics[width=\linewidth]{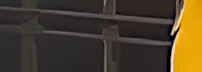} \\
       \includegraphics[width=\linewidth]{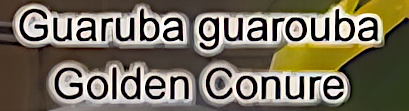}  
  \caption{SRWD-VSR}
\end{subfigure}
\caption{\textbf{Qualitative Comparison for all ablations.} These zoomed-in scenes are two times super-resolved images using all the versions of the proposed SRWD-VSR. It can be noticed that the overall results of SRWD-VSR, i.e., (e), (j), and (o), are visibly more plausible. the letter `a' in (j), all the text in (o), and the eye of the bird with much fewer artifacts. \textbf{(Zoom-in for best view)}}
\label{qual_self}
\end{figure*}

%% file: tables/quant_self_degra.tex
\begin{table}[t]
\centering
\renewcommand{\arraystretch}{1.1}
\setlength{\tabcolsep}{2pt}
\caption{\textbf{Degradation analysis of our VSR algorithm.} Symbols `$\uparrow$' indicates a higher value is better while `$\downarrow$' indicates a lower value
is better. Overall, SRWD-VSR performs consistently better. (Note: These results are for $\times$ 2 super-resolution)}
\begin{tabular}{|C{.3cm}|C{2.7cm}||C{1.4cm}|C{1.7cm}|}
\hline
\textbf{} & \textbf{Algorithm} &  \textbf{NRQM$\uparrow$}  & \textbf{BRISQUE$\downarrow$}  \\ 
\hline \hline
\multirow{4}{*}{\rotatebox[origin=c]{90}{K$\mid$Lens}} & SRWD-VSR  & \textbf{7.1013} & \textbf{26.9868}  \\

& SRWD-VSR-KG  & 6.9201 &  27.0948 \\

& SRWD-VSR-SN & 6.8014 & 27.1512  \\


& SRWD-VSR-PB  & 6.8812 & 27.2002  \\ \hline

\hline
\multirow{4}{*}{\rotatebox[origin=c]{90}{RealVSR}} & SRWD-VSR & \textbf{6.7507} & \textbf{29.1829}  \\

& SRWD-VSR-KG  & 6.5480 & 29.9901  \\

& SRWD-VSR-SN & 6.5210 & 30.0124  \\


& SRWD-VSR-PB  & 6.5912 & 29.9189  \\ \hline

\hline
\multirow{4}{*}{\rotatebox[origin=c]{90}{VideoLQ}} & SRWD-VSR & \textbf{6.9054} &  \textbf{29.4897} \\

& SRWD-VSR-KG & 6.8112 & 29.6814  \\

& SRWD-VSR-SN & 6.7614 & 29.6610  \\

& SRWD-VSR-PB  & 6.7514 & 29.6150  \\ \hline
\end{tabular}
\label{qual_self_table_deg}
\end{table} 

%% file: figures/qual_sota.tex
\begin{figure*}
\begin{subfigure}{.3\textwidth}
  \centering
  \includegraphics[width=0.8\linewidth, height=30mm]{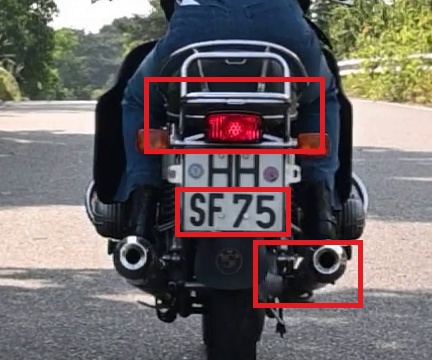}  
  \caption{Our Dataset}
\end{subfigure}
\begin{subfigure}{.13\textwidth}
  \centering
      \includegraphics[width=\linewidth]{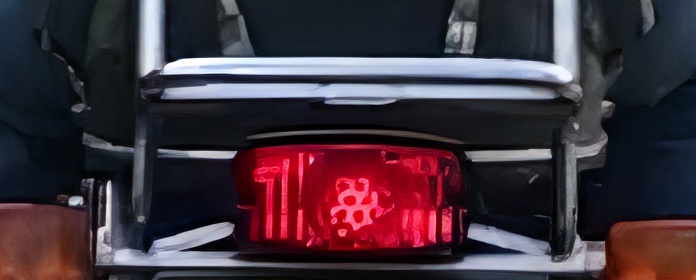}  \\
  \includegraphics[width=\linewidth]{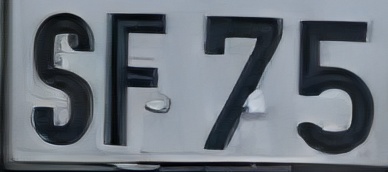}  \\
    \includegraphics[width=\linewidth]{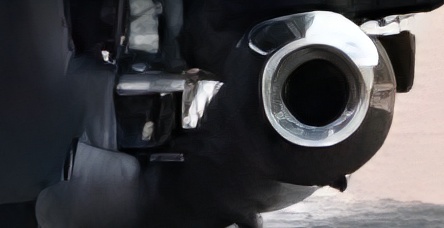}  
  \caption{BSRGAN\cite{bsrgan}}
\end{subfigure}
\begin{subfigure}{.13\textwidth}
  \centering
      \includegraphics[width=\linewidth]{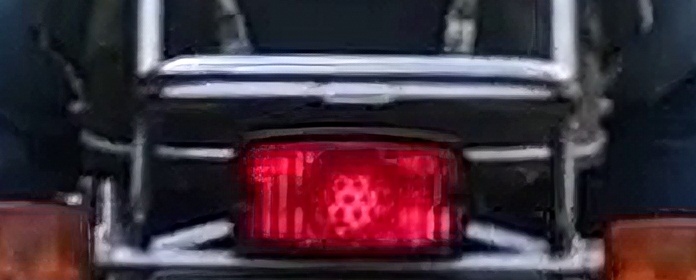} \\
  \includegraphics[width=\linewidth]{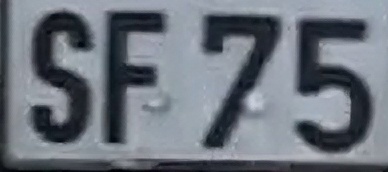}  \\
    \includegraphics[width=\linewidth]{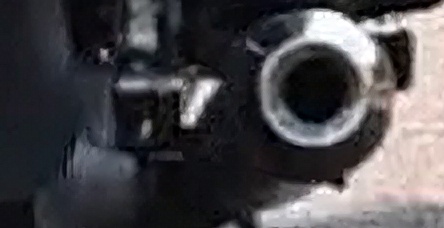}  
  \caption{TecoGAN\cite{tecoGAN}}
\end{subfigure}
\begin{subfigure}{.13\textwidth}
  \centering
      \includegraphics[width=\linewidth]{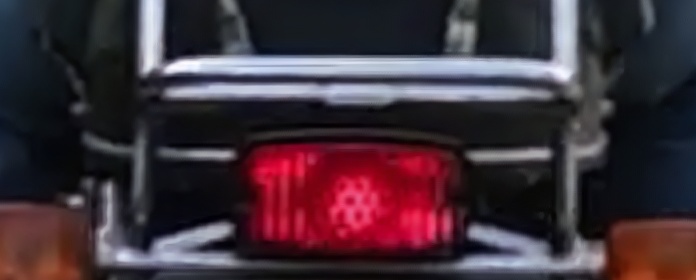} \\
  \includegraphics[width=\linewidth]{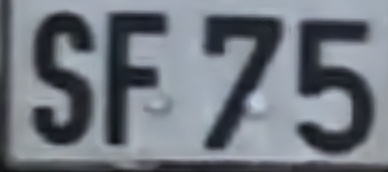}  \\
    \includegraphics[width=\linewidth]{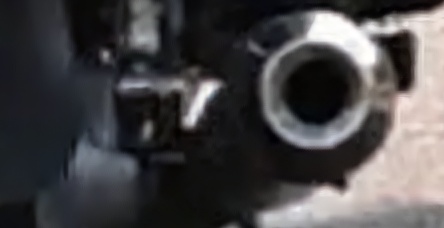}  
  \caption{RealVSR\cite{real_2021}}
\end{subfigure}
\begin{subfigure}{.13\textwidth}
  \centering
 \includegraphics[width=\linewidth]{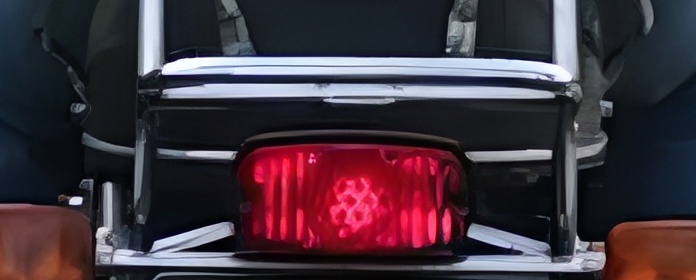} \\
  \includegraphics[width=\linewidth]{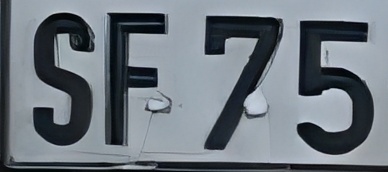}  \\
    \includegraphics[width=\linewidth]{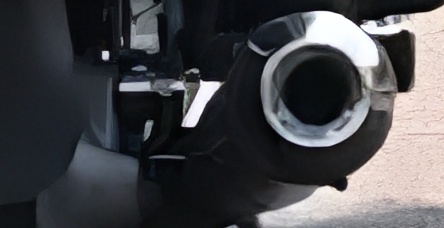}  
  \caption{BasicVSR\cite{real_basic_2022}}
\end{subfigure}
\begin{subfigure}{.13\textwidth}
  \centering 
      \includegraphics[width=\linewidth]{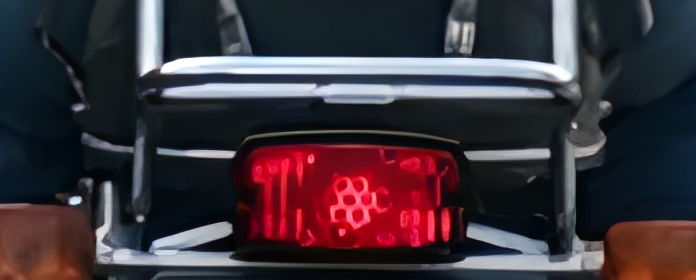} \\
  \includegraphics[width=\linewidth]{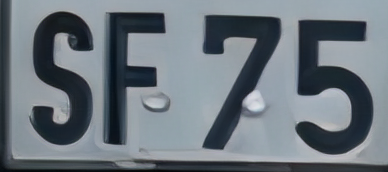}  \\
    \includegraphics[width=\linewidth]{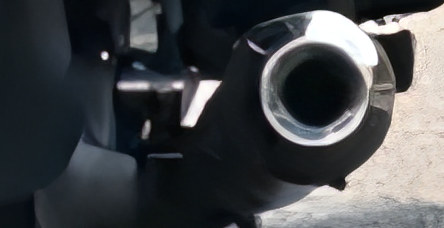} 
  \caption{SRWD-VSR}
\end{subfigure}
\newline
\begin{subfigure}{.3\textwidth}
  \centering
  \includegraphics[width=.8\linewidth, height=30mm]{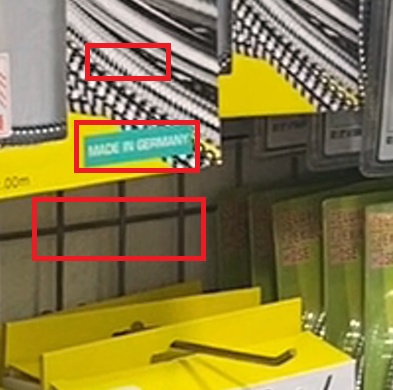}  \caption{RealVSR\cite{real_2021}}
\end{subfigure}
\begin{subfigure}{.13\textwidth}
  \centering
    \includegraphics[width=\linewidth]{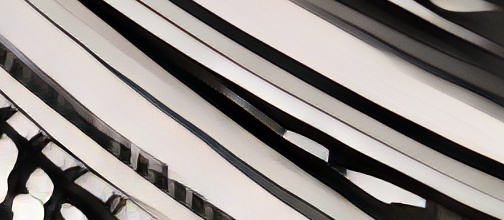} \\
  \includegraphics[width=\linewidth]{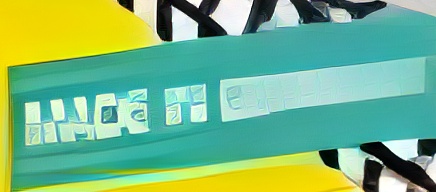}  \\
    \includegraphics[width=\linewidth]{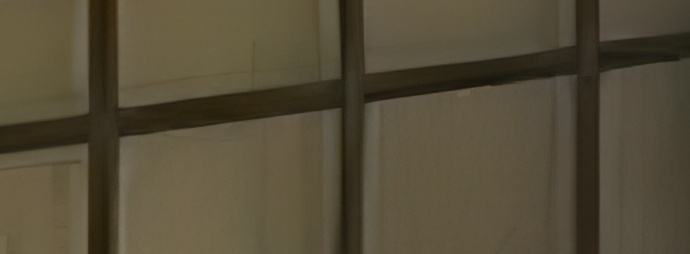}  
  \caption{BSRGAN\cite{bsrgan}}
\end{subfigure}
\begin{subfigure}{.13\textwidth}
  \centering
      \includegraphics[width=\linewidth]{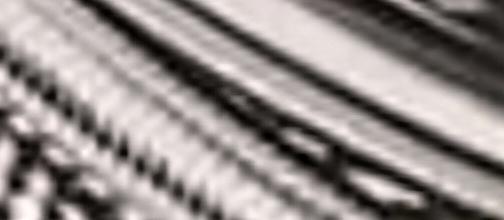} \\
  \includegraphics[width=\linewidth]{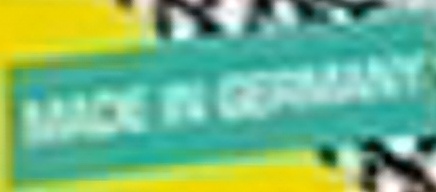}  \\
    \includegraphics[width=\linewidth]{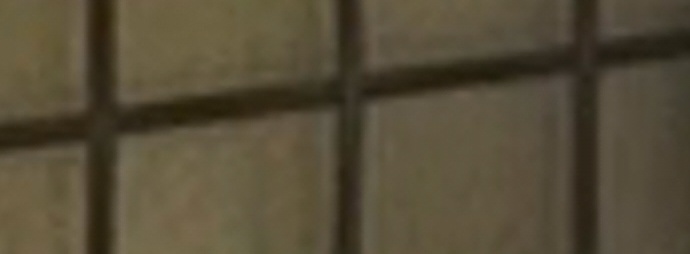}  
  \caption{TecoGAN\cite{tecoGAN}}
\end{subfigure}
\begin{subfigure}{.13\textwidth}
  \centering
       \includegraphics[width=\linewidth]{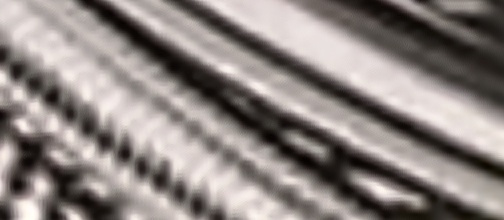} \\
  \includegraphics[width=\linewidth]{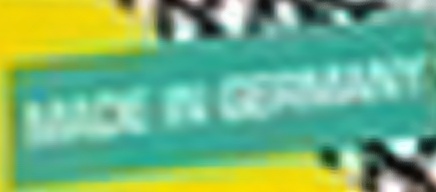}  \\
    \includegraphics[width=\linewidth]{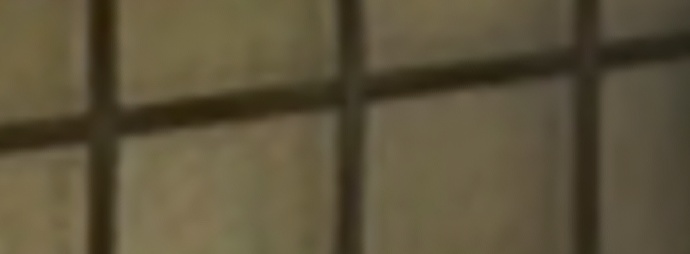} 
  \caption{RealVSR\cite{real_2021}}
\end{subfigure}
\begin{subfigure}{.13\textwidth}
  \centering
      \includegraphics[width=\linewidth]{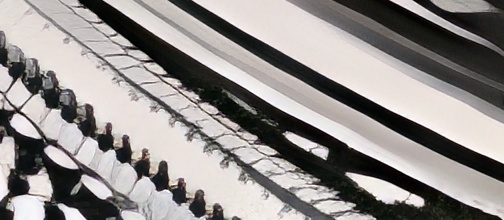} \\
  \includegraphics[width=\linewidth]{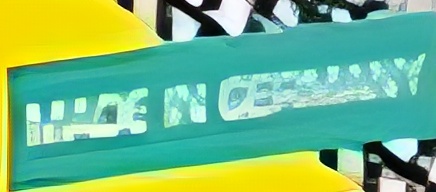}  \\
    \includegraphics[width=\linewidth]{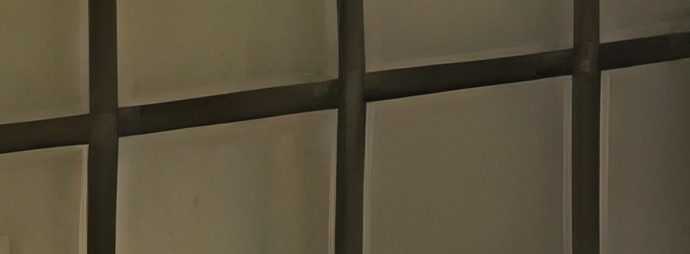}  
  \caption{BasicVSR\cite{real_basic_2022}}
\end{subfigure}
\begin{subfigure}{.13\textwidth}
  \centering    
  \includegraphics[width=\linewidth]{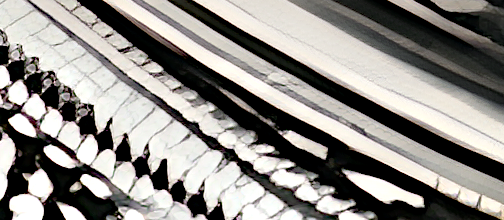} \\
  \includegraphics[width=\linewidth]{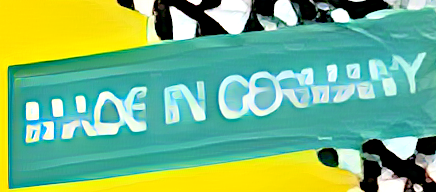}  \\
    \includegraphics[width=\linewidth]{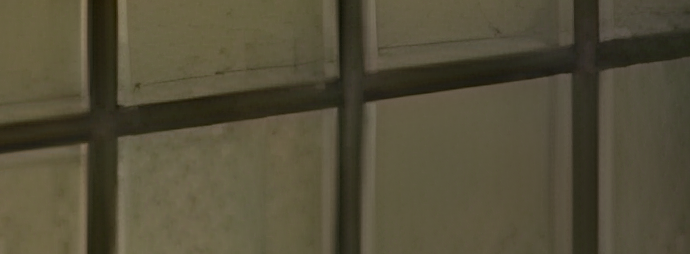} 
  \caption{SRWD-VSR}
\end{subfigure}
\newline
\begin{subfigure}{.3\textwidth}
  \centering
  \includegraphics[width=.8\linewidth, height=30mm]{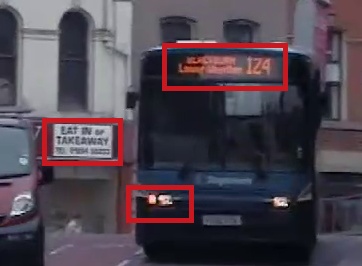}  
  \caption{VideoLQ\cite{real_basic_2022}}
\end{subfigure}
\begin{subfigure}{.13\textwidth}
  \centering
    \includegraphics[width=\linewidth]{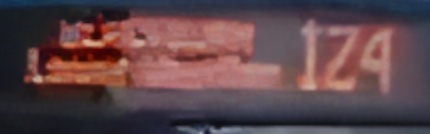}  \\
      \includegraphics[width=\linewidth]{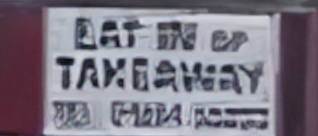}  \\
    \includegraphics[width=\linewidth]{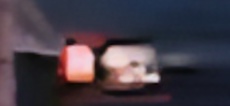} 
  \caption{BSRGAN\cite{bsrgan}}
\end{subfigure}
\begin{subfigure}{.13\textwidth}
  \centering
    \includegraphics[width=\linewidth]{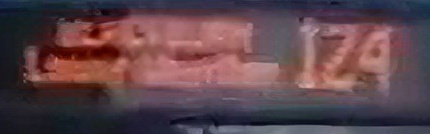}  \\
      \includegraphics[width=\linewidth]{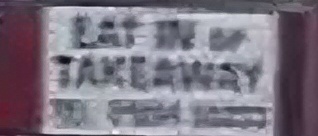}  \\
    \includegraphics[width=\linewidth]{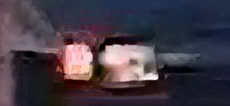} 
  \caption{TecoGAN\cite{tecoGAN}}
\end{subfigure}
\begin{subfigure}{.13\textwidth}
  \centering
    \includegraphics[width=\linewidth]{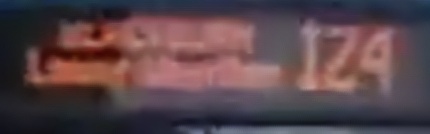}  \\
      \includegraphics[width=\linewidth]{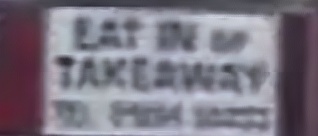}  \\
     \includegraphics[width=\linewidth]{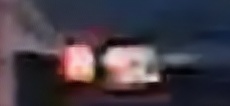}
  \caption{RealVSR\cite{real_2021}}
\end{subfigure}
\begin{subfigure}{.13\textwidth}
  \centering
    \includegraphics[width=\linewidth]{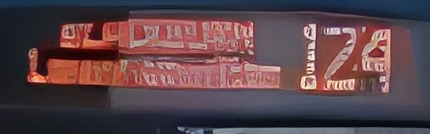}  \\
      \includegraphics[width=\linewidth]{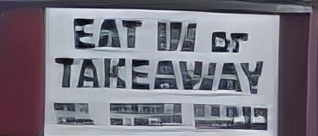}  \\
     \includegraphics[width=\linewidth]{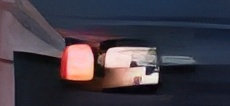} 
  \caption{BasicVSR\cite{real_basic_2022}}
\end{subfigure}
\begin{subfigure}{.13\textwidth}
  \centering
    \includegraphics[width=\linewidth]{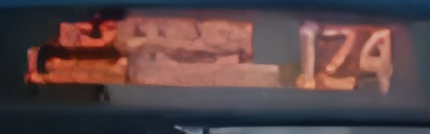} \\
      \includegraphics[width=\linewidth]{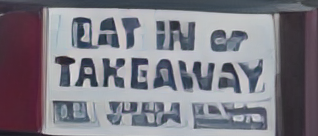}  \\
     \includegraphics[width=\linewidth]{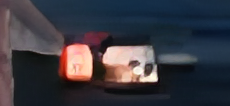} 
  \caption{SRWD-VSR}
\end{subfigure}
\caption{\textbf{Comparison with SR Algorithms.} We compared SRWD-VSR with four SOTA algorithms i.e., \cite{bsrgan, tecoGAN, real_2021, real_basic_2022} ($\times$ 4 SR). Overall, the SRWD-VSR performs better visibly, for example, `Made in Germany' text in (l) and `124' in (r). \textbf{(Zoom in for best view)}}
\label{qual_sota_fig}
\end{figure*}

%% file: tables/quant_sota.tex
\begin{table}[t]
\centering
\renewcommand{\arraystretch}{1.02}
\setlength{\tabcolsep}{0pt}
\caption{ \textbf{Performance comparison of proposed SRWD-VSR with existing algorithms.} `-' symbol indicates the unavailability of the pre-trained weights or the training script for that method. It can be observed that the proposed SRWD-VSR performs better in terms of NRQM and BRISQUE parameter values.}
\begin{tabular}{|c|c||C{1.5cm}|C{1.7cm}|C{1.5cm}|C{1.7cm}|}\hline
\multirow{2}{*}{\textbf{}} & \multirow{2}{*}{\textbf{}} & \multicolumn{2}{c|}{\textbf{ $\times$2 SR}} & \multicolumn{2}{c|}{\textbf{$\times$4 SR}}  \\
\cline{3-6}
& & NRQM$\uparrow$   & BRISQUE$\downarrow$ & NRQM$\uparrow$ &  BRISQUE$\downarrow$  \\ 
\hline \hline

\multirow{6}{*}{\rotatebox[origin=c]{90}{RealVSR\cite{real_2021}}} & Proposed  & \textbf{6.7507} &  \textbf{29.1829}  & \textbf{6.7257}  &  \textbf{17.5603} \\

& RBVSR\cite{real_basic_2022} & - & - & 6.4080 &   23.6932 \\

& RVSR\cite{real_2021} & 3.6927  & 46.6572  &  2.8689 &   51.7278 \\

& Teco\cite{tecoGAN} & 6.7251  & 38.1057 & 3.0610 &  50.1514  \\

& BSR\cite{tecoGAN} & 5.1487  & 37.3136 & 5.7911 &  26.3568 \\

& RESR\cite{realesrgan} & 5.8664 &  30.4356 & 6.0012 &  28.1113 \\

\hline
\multirow{6}{*}{\rotatebox[origin=c]{90}{VideoLQ\cite{real_basic_2022}}} & Proposed & \textbf{6.9054} &   \textbf{29.4897} & \textbf{6.5993} &   \textbf{23.3656} \\

& RBVSR\cite{real_basic_2022} & - & - & 6.1618 &  27.5545 \\

& RVSR\cite{real_2021}  & 3.3715 & 31.9713 & 3.0727 &  43.7101 \\

& Teco\cite{tecoGAN} & 6.1405 & 30.9813 & 5.2951 &  32.7901  \\

& BSR\cite{bsrgan} & 6.3892  & 30.4867 & 6.3858 &  30.4369 \\

& RESR\cite{realesrgan} & 6.3690  & 27.7421 & 5.8140 &  30.5699 \\
\hline

\multirow{5}{*}{\rotatebox[origin=c]{90}{K$|$Lens Dataset}} & Proposed  & \textbf{7.1013} & \textbf{26.9868} & \textbf{6.4678} & \textbf{24.5906} \\

& RBVSR\cite{real_basic_2022} & - & - &  6.1259 &  24.9979  \\

& RVSR\cite{real_2021} & 3.3510   & 40.5177 & 5.7346 &  47.9872  \\


& BSR\cite{bsrgan} & 5.3301  & 33.3556 & 5.1832 & 33.0617 \\

& RESR\cite{realesrgan} & 5.0024  & 34.3215 & 4.8346 &  35.5532 \\

\hline
\end{tabular}
\label{qual_sota_table}
\end{table}